\documentclass[12pt]{article}
\usepackage{amssymb}
\usepackage[dvips]{graphicx,color}
\usepackage{wrapfig}
\usepackage{makeidx}

\usepackage{ulem}	
\makeatletter
  \renewcommand{\theequation}{%
     \thesection.\arabic{equation}}
  \@addtoreset{equation}{section}
\makeatother
\usepackage{amsmath}

\def\bfr{\begin{flushright}}
\def\efr{\end{flushright}}
\def\si{\quad}
\def\sii{\qquad}

\def\noi{\noindent}
\def\vs{\vspace}
\def\vss{\vspace{.5cm}}
\def\begc{\begin{center}}
\def\endc{\end{center}}
\def\be{\begin{eqnarray}}
\def\ee{\end{eqnarray}}
\def\eq{\label}
\def\nn{\nonumber  \\}
\def\qed{$\blacksquare$}
\def\rarwww#1{\smash{%
             \mathop{\hbox to 1.3cm{\rightarrowfill}}
             \limits^{#1}}}
\def\rarw{\rightarrow}

\def\und{\underline}
 
\def\a{\alpha}
\def\b{\beta}

\def\d{\delta}
\def\e{\epsilon}

\def\h{\eta}
\def\th{\theta}

\def\k{\kappa}
\def\l{\lambda}
\def\L{\Lambda}
\def\m{\mu}
\def\n{\nu}
\def\x{\xi}

\def\o{\o}
\def\p{\pi}
\def\r{\rho}
\def\s{\sigma}
\def\t{\tau}

\def\f{\phi}
\def\c{\chi}
\def\ps{\psi}

\def\ve{\varepsilon}

\def\vf{\varphi}

\def\Ps{\Psi}

\def\dal{\rule[-1mm]{0.1mm}{4mm}
         \rule[3mm]{4mm}{0.1mm} 
         \hspace{-4mm}
         \rule[-1mm]{4mm}{0.1mm} 
         \rule[-1mm]{0.1mm}{4mm}
         \hspace{1mm}}

\def\bra{\langle}
\def\ket{\rangle}
\def\fr{\frac}
\def\del{\partial}

\def\mxxb{\left( \begin{array}{cc}}           
\def\mxxe{\end{array} \right)}
\def\mxxxb{\left( \begin{array}{ccc}}         
\def\mxxxe{\end{array} \right)}
\def\mxxxxb{\left( \begin{array}{cccc}}       
\def\mxxxxe{\end{array} \right)}
\def\mxxxxxb{\left( \begin{array}{ccccc}}     
\def\mxxxxxe{\end{array} \right)}
\def\mxxxxxxb{\left( \begin{array}{cccccc}}     
\def\mxxxxxxe{\end{array} \right)}
\def\mxxxxxxxb{\left( \begin{array}{ccccccc}}     
\def\mxxxxxxxe{\end{array} \right)}               

\def\kakkob{\left\{ \begin{array}{c}}
\def\kakkoe{\end{array}  \right. }
\def\vecb{\left( \begin{array}{c}}
\def\vece{\end{array} \right) }

\def\bm#1{\mbox{\boldmath #1}}

\def\inte{\Bbb Z}

\def\real{\Bbb R}

\def\bm#1{\mbox{\boldmath #1}}

\def\bbC{\bm{$C$}}

\def\bbP{\bm{$P$}}

\def\bbU{\bm{$U$}}

\def\bbW{\bm{$W$}}

\def\bba{\bm{$a$}}
\def\bbb{\bm{$b$}}

\def\bbf{\bm{$f$}}

\def\bbk{\bm{$k$}}

\def\bbp{\bm{$p$}}

\def\bbu{\bm{$u$}}

\def\bbw{\bm{$w$}}
\def\bbx{\bm{$x$}}
\def\bby{\bm{$y$}}
\def\bbz{\bm{$z$}}

\def\bbpi{\bm{$\p$}}

\def\bb1{\bm{$1$}}
\def\bbzero{\bm{$0$}}

\def\bbdel{\bm{$\del$}}

\newfont{\bg}{cmr10 scaled\magstep3}

\def\sen{\rm -----------------------------------------------------------------------------------}

\def\deff{\stackrel{\rm  def}{=}}

\newtheorem{xx}{}[section]
\def\pdel#1#2
{
\fr{\del {#1}}{\del {#2}}
}
\def\ppdel#1#2#3
{
\fr{\del^2 {#1}}{\del {#2}\del {#3}}
}

\newcommand{\sageru}[1]{\smash{\lower2.0ex\hbox{#1}}}

\newfont{\bbbg}{cmr10 scaled\magstep2}

\newcommand{\bbzerou}{\smash{\lower1.7ex\hbox{\bbzero}}}

\def\rkakkob{\left. \begin{array}{c}}
\def\rkakkoe{\end{array}  \right\} }

\def\vDal{\varDelta}

\begin{document}
\bfr
DIP 2406-01\\
\efr

\vs{0.3cm}

\begc
{\LARGE{\bf  Theory of Complex Particle without 
Extra Dimensions}}
 \vss

Takayuki Hori
\footnote{email: hori@tokyo.zaq.jp}\vs{0.3cm}

Doyo-kai Institute of Physics, 3-46-4,
Kitanodai, \\
Hachiouji-shi, Tokyo 192-0913, Japan
\endc

\vs{0.3cm}

 \vss

\begin{abstract}
Complex particle is a kind of bilocal particle having unexpected symmetry, which was proposed by the authour. In the present paper, we show that critical dimension of the complex particle in Minkowski spacetime is 
$D = 4$, while $D = 2, 4$ or $6$ are permitted in Euclid spacetime. The origin of the restriction to the dimension is the existence of tertiary constraint in the canonical theory, quantization of which leads to an eigenvalue equation having single-valued and bounded solutions only in particular dimension of spacetime. The derivation is based on a detailed analysis of Laplace-Beltrami operator on 
$S^{1,D-2}$ or $S^{D-1}$. 
\end{abstract}

\setlength{\baselineskip}{15pt}
\setcounter{page}{1}
\pagenumbering{arabic}

\section{Introduction}

In the past two decades the investigations of  string theorists may be devoted to the ploblem of the vacuum structure of the world, called landscape \cite{kklt},\cite{susskind},  
which is consistent to the low energy phenomenology.
The motivation for that work is based on the premise that the extra dimensions of spacetime exist, which is far from the present experimental evidence.
On the other hand, no realistic model of particle theory has been proposed, whose critical dimension equals to four.

In the present paper we show that there is a model  without  extra dimensions, though it is a toy model at present.
The key element leading to the restriction to the dimension is the tertiary constraint which exists, in general, in such a model that the  number of gauge degrees of freedom is less than that of the first class constraints in the canonical theory, {\it i.e.},
the breakdown of Dirac's conjecture 
\cite{dirac_1}-\cite{rothe}.

The model is complex particle proposed by the author, and a preliminary study has been reported  \cite{hori_04},\cite{hori_05}.
The dynamical variables of the model are complex  coordinates of an object in $D$-dimensional flat spacetime,
the real and the imaginary parts of which may be
representing two relativistic particles.
However, the two coordinates are not independent on each other due to a bilinear {\it interaction term}
in the lagrangian.
In that sense the complex particle is resembling to the bilocal particle \cite{hori_01}-\cite{hori_03} which has
unexpected global symmetry.
The lagrangians of these models have gauge symmetry  generalising the reparametrization invariance of the relativistic particle, and have the corresponding BRST symmetry. 
One unsolved problem in the bilocal particle model is that the ghost numbers, associated with the BRST  symmetry, of the states in the physical space are not zero \cite{hori_03}, and the ordinary quantization scheme may not be applicable.
Fortunately the complex particle has not the problem of the quantization  appeared in the bilocal particle.

In the quantum theory 
the tertiary constraint in the complex particle 
is replaced by differential equations.
In solving them we get, among other equations, an eigenvalue equation of d'Alembert operator, and solving it  results in the eigenvalue problem of the Laplace-Beltrami(LB) operator  on $S^{D-1}$ and  $S^{1,D-2}$, corresponding to the Euclid and the Minkowski metric, respectively.

The LB operator in $D=3$ is the square of the angular momentum operator, and the eigenvalues are $\ell(\ell + 1),~(\ell\in\inte_{\ge 0}$) as is well known in the elementary course of quantum mechanics.
The eigenvalue problem of LB in $D$-dimensional Euclid space has been well investigated in the last century, published in a mathematical literature in the context of the representation theory of the orthogonal groups \cite{dai}-\cite{nomura}.
However, the eigenvalue problem in the Minkowski  spacetime has not been well studied because of the less application to physics.
We find some papers in this era \cite{strichartz_2},\cite{limic}.

In the present paper we find the complete solution to
 the eigenvalue problem of LB in the $D$-dimensional Minkowski or Euclid spacetime, with some physical conditions such as single-valuedness or boudedness.
On the basis of this result we conclude that the dimension of spacetime being allowed in the complex particle in the Minkowski spacetime is 4, 
and $D=2,4$ or $6$ in the Euclid spacetime.\vss

In Section 2 and 3 we give an outline of the classical theory of the complex particle.
In Section 4 we give the quantization of the model and the equations representing the physical state condition. 
In Section 5 the mathematical aspect of LB operator is illustrated.
Section 6 is devoted to derivation of the critical dimension of spacetime.
Finally in Section 7 we give some speculations for future study of the present model.

\section{Lagrangian of complex particle}

Let us start with a system with complex coordinates, $\bbz =(z^\m),~(0 \le \m \le D-1)$, which describes 
an object moving in $D$-dimensional complex manifold.
The trajectory of it is parametrized by {\it time 
 variable}, $\t$.
Introducing the complex einbein, $g$, 
we can write the action as
\be
I = \int d\t~ L,
\sii
L
=
\fr{\bbw^2}{2g} + i\k\bbw\cdot\bar{\bbz} + {\rm c.c.},
\sii
\bbw = \dot{\bbz},
\ee
where c.c. is the complex conjugate and $\k$ is a constant with dimension of mass square
\footnote[2]
{Here and hereafter we use the notation for $D$-dimensional vectors $\bba,~\bbb$, as
$\bba\cdot\bbb=a_\m b^\m$.
The flat metric of spacetime is $\h^{\m\n}={\rm diag}(\pm 1,1,..,1)$, where the $\pm$ sign is $+1$ and $-1$ for Euclid and Minkowski spacetime, respectively.}.
The real and the imaginary parts of $\bbz$ describe
 real coordinates of two relativistic particles traveling in our $D$-dimensional spacetime manifold.
However, we call the system a single complex particle
 instead of two particles, because the Poisson bracket between the two momenta does not vanish, as is shown later, and they have not simultaneous eigenvalues in quantum theory.
Actually one of the two particles has wrong sign in the kinetic term, {\it i.e.}, behaves as ghost.  

 \vss

The Euler-Lagrange(EL) equation obtained by minimizing the action under the variation of $\bbz$ is
\be
{\bf [EL]} &\deff&
\fr{d}{d\t}
\left(
\fr{\bbw}{g} + 2i\k\bar{\bbz}
\right)
= 0.
\sii ({\rm and}~{\rm c.c.})
\eq{CPLX_EL}
\ee
Another equation is obtained by varying 
the einbein $g$:
\be
\ell &\deff&
-g^2{\bf [EL]}_g = \bbw\cdot\bbw = 0.
\sii ({\rm and}~{\rm c.c.})
\eq{CPLX_LagC}
\ee
The dynamical variables are the coordinate variables, $(\bbz,g)$
and their velocities, $(\bbw,w=\dot{g})$, 
(and their complex conjugates).
The Euler-Lagrange equations which are first rank differential equations with respect to $\t$, determine the time development of the dynamical variables under some initial value conditions.
In that sense Eq.(\ref{CPLX_LagC})
is not an equation of motion determining time-development but a constraint
equation for the dynamical variables.
Such a constraint is called lagrangian constraints \cite{kamimura}. 
The lagrangian constraint should  be preserved  not only at initial time but along the trajectory,. 
Since the time derivative of $\ell$
is
\be
\dot{\ell}
=
\fr{2w}{g}\ell - 4i\k g \bbw\cdot\bar{\bbw},
\ee
we have the condition
\be
\ell_0 \deff \bbw\cdot\bar{\bbw} = 0,
\eq{CPLX_2ndLC}
\ee
provided $\ell=0$.
The time preservation condition of a lagrangian constraint is called 2nd order lagrangian constraint, 
which should be also preserved, yielding 3rd order lagrangian constraint, and repeated on.
In the present case the lagrangian constraint is over at the 2nd order.
It is worth while to note that the 2nd order lagrangian constraint (\ref{CPLX_2ndLC}) 
exists only if $\k \ne 0$.

The 1st integral of (\ref{CPLX_EL}) 
is
\be
\bbw = -2ig\k(\bar{\bbz} - \bar{\bbC})
\ee	
where $\bbC$ is an arbitrary constant vector.
Differentiating the above equation and $\bbw = \dot{\bbz}$,
we get the second order differential equation,
\be
\ddot{y} - \r\dot{y} - \h y = 0,
\eq{CPLX_ddy+dy+y=}
\ee
where we put $\bbz - \bbC \rarw y,~\r=w/g,~\h=4\k^2g\bar{g}$.
The solution for $\bbz$ is
\be
\bbz(\t) = \bbC + \bbz_{(0)}f(\t)
\eq{CPLX_z(t)=}
\ee
where $y=f(\t)$ is a solution to (\ref{CPLX_ddy+dy+y=}) and $\bbz_{(0)}$ is arbitrary constant vector.
It is not necessary to solve (\ref{CPLX_ddy+dy+y=})
 generally in order to get expression of the trajectory in the target space as is shown below.

From the 0-th component of
(\ref{CPLX_z(t)=})
we have
\be
f(\t) =  \fr{1}{z_{(0)}^0}(z^0(\t) - C^0),
\ee
and substituting it to the spacial component of
(\ref{CPLX_z(t)=})
we get
\be
z^i(\t)
&=&
c^i + v^iz^0(\t),
\eq{CPLX_z^i=}
\ee
where
\be
c^i = \fr{C^{[i}z_{(0)}^{0]}}{z_{(0)}^0},
\sii
v^i = \fr{z_{(0)}^i}{z_{(0)}^0},
\sii
(i=1,..,D-1).
\eq{CPLX_vi=}
\ee

Now let us consider the physical content of the above solution in the Minkowski spacetime ($\h_{00}= -1$).
With respect to the time component of $\bbz$, 
one must set a relation between the {\it time}, 
$z^0$, of the target space  and the internal time, $\t$.
We set
\be
\Re z^0(\t) = t,
\sii
\Im z^0(\t) = 0.
\eq{CPLX_Rz0=t_Iz0=0}
\ee
Then the real and imaginary parts of
(\ref{CPLX_z^i=}),
$z^i=c^i + v^it$,
represent a particle moving in the target space 
with constant {\it velocities}, $\Re v^i$ and $\Im v^i$, respectively, with respect to the time $t$.

The condition (\ref{CPLX_Rz0=t_Iz0=0}) should be used after solving the lagrangian constraints.
Since 
(\ref{CPLX_z(t)=})
holds on the trajectory satisfying EL equation,
we have 
$
\bbw = \bbz_{(0)}\dot{f}(\t),
$
on the trajectory.
Then
the 1st order lagrangian constraint
becomes
$
(\bbz_{(0)})^2=0
$.
Hence from the second equation of 
(\ref{CPLX_vi=}) we have 
\be
\sum_{i=1}^{D-1}v^iv_i = 1,
\ee
here we use $\h_{00}= - 1$.
Putting
$v_1^i = \Re v^i,~v_2^i = \Im v^i$,
this condition becomes
\be
\sum_{i=1}^{D-1}(v_1^iv_{1i} - v_2^iv_{2i}) = 1,
\sii
\sum_{i=1}^{D-1}v_1^i v_{2i} =  0.
\eq{CPLX_v1v2_1}
\ee
This means that the real part of the complex coordinates represent a particle moving faster than light, {\it i.e.} a tachyon, while the imaginary part represent one moving perpendicular to the former.

The 2nd order lagrangian constraint
(\ref{CPLX_2ndLC}),
which exists if $\k \ne 0$,
changes the above picture drastically.
The similar manipulation as that of the 1st order lagrangian constraint gives 
\be
\sum_{i=1}^{D-1}(v_1^iv_{1i} + v_2^iv_{2i})=  1,
\eq{CPLX_v1v2_2}
\ee
and, in conjunction with (\ref{CPLX_v1v2_1}),
we conclude
\be
\sum_{i=1}^{D-1}v_1^iv_{1i}  = 1,
\sii
v_2^i = 0.
\ee
That is, the real part of $\bbz$ represent a moving particle in a constant direction with the velocity of light, and the imaginary part of $\bbz$ represent a particle staying at rest. 
In another words the latter may be interpreted as a kind of conserved quantity.
The coordinates of the complex particle in the Minkowski spacetime may be  viewed as composite of the ordinary coordinates of particle moving with velocity of light and a conserved quantity, the conservation of the latter is guaranteed by the EL equations.
In the above discussion we see the vanishing limit of $\k$ does not coincide with the system which has $\k = 0$ from outset.
This situation is seen in many respects of the complex particle.

Further insight is obtained by dividing lagrangian 
in terms of the real and imaginary parts of $\bbw$ and $w$  (hereafter we consider both of the Euclid and Minkowski signature).
The kinetic terms are expressed as
\be
L_{\rm kin}
=
\fr{2}{g_1^2 + g_2^2}
[
g_1(\bbu^2 - \bbf^2)
+
2g_2\bbu\cdot\bbf
],
\ee
where  $\bbw = \bbu + i\bbf$ and $g = g_1 + ig_2$.
If we put conventionally $g_1>0$,
the signs of the kinetic terms are plus for the real part of $z$ and minus for the imaginary part.
Hence the imaginary part of the complex coordinates are the ghost variables.\vss

Thus the classical behaviour of the complex particle is restricted in a somehow  curious manner. This is a consequence of existence of unexpected symmetry of the action.
Generalizing the reparametrization invariance of 
the action of two relativistic particles,
we consider the following transformation of $\bbz$ and $g$:
\be
\d \bbz = \ve \bbw + \fr{\ve_0}{\bar{g}}\bar{\bbw},
\sii
\d g = \x,
\eq{CPLX_d_sg}
\ee
and c.c. of them,  
where $\ve,~\ve_0$ and $\x$ are $\t$-dependent infinitesimal 
parameters, and $\ve_0$ is a real parameter.
After straightforward but rather lengthy calculation 
(see {\bf Appendix \ref{App_dL_xx}}),
we see the variation of the lagrangian under the transformation (\ref{CPLX_d_sg}) is
\be
\d L
&=&
\fr{\ell_0}{2g\bar{g}}
[\dot{\ve}_0 - 2i\k g\bar{g}(\ve - \bar{\ve})]
+
A\ell + \fr{dE}{d\t}
+
{\rm c.c.},
\eq{CPLX_dL_cplx}
\ee
where
\be
A 
\deff
\fr{1}{2g^2}
\left(
\fr{d}{d\t}(\ve g) + 4i\k g\ve_0 - \x
\right),
\sii
E
\deff
\left(
\fr{\bbw}{2g} + i\k\bar{\bbz}
\right)\cdot\d\bbz.
\eq{CPLX_def_A_E}
\ee
The transformation under which the action is invariant apart from the lagrangian constraints
is called semi-gauge transformation \cite{hori_1902}.
Hence  (\ref{CPLX_d_sg}) is a semi-gauge transformation for arbitrary infinitesimal parameters, $\ve,\ve_0$ and $\x$.

In particular if we choose the parameters so that
\be
&&
\dot{\ve}_0 - 2i\k g\bar{g}(\ve - \bar{\ve}) = 0,
\eq{CPLX_dot_e_0=}
\\
&&
\x =  \fr{d}{d\t}(\ve g) + 4i\k g\ve_0,
\eq{CPLX_gzi=}
\ee
then the transformation is a gauge transformation,
{\it i.e.}, the action is invariant under it.
Hereafter we set $\x$ as (\ref{CPLX_gzi=}) so that $A=0$.
It is important to note that 
(\ref{CPLX_dot_e_0=})
determines $\ve_0$, except its  constant mode,  completely by $\ve,~\bar{\ve}$.
In that sense the number of the gauge degrees of freedom of the complex particle is essentially two.
The global invariance under the transformation generated by the constant mode of $\ve_0$ is a hidden one, which is in fact a generalization of what exists in the system of free relativistic two particles.

What is the conservation law derived from the above global symmetry? 
We see from (\ref{CPLX_dL_cplx}) that the lagrangian varies under the global transformation as 
\be
\d L
=
\fr{d}{d\t}
\left[
\fr{\ve_0\bar{\bbw}}{\bar{g}}
\cdot
\left(
\fr{\bbw}{2g}
+
i\k\bar{\bbz}
\right)
\right]
+ {\rm c.c.}
\eq{CPLX_dL(const.dz)}
\ee
On the other hand we have,
up to EL-equations,
\be
\d L
=
\fr{d}{d\t}
\left(
\fr{\ve_0\bar{\bbw}}{\bar{g}}
\cdot
\fr{\del L}{\del \bbw}
\right)
+ {\rm c.c.}.
\ee
Equating them we see that the quantity
\be
Q
=
\fr{\bar{\bbw}}{\bar{g}}
\cdot
\left(
\fr{\bbw}{2g}
+
i\k\bar{\bbz}
-
\fr{\del L}{\del \bbw}
\right)
=
\fr{\bar{\bbw}\cdot\bbw}{2\bar{g}g}
\ee
is conserved along the trajectory satisfying the EL-equations.
However, $Q$ vanishes by the 2nd order lagrangian constraint.
Thus we get no nontrivial conserved quantity from the global invariance.

Another conserved quantity is the $D$-dimensional momentum
\be
\bbp \deff 
\fr{\bbw}{g} + 2i\k\bar{\bbz},
\eq{CPLX_consv_p}
\ee
derived from the translation invariance of the action.

 \vss

\section{Canonical theory}
 
Now, let us proceed to the canonical theory.
Denote the canonical momenta of $\bbz$ and the complex einbein $g$ as $\bbpi$ and $\p$, respectively.
The characteristic equations $\bbpi=\del L/\del \bbw$ and $ p = \del L/\del w$
are written as
\be
\bbpi = \fr{\bbw}{g} + i\k\bar{z},
\sii
\p = 0.
\eq{CPLX_consv_p-pi}
\ee
The general solution, $\bbw = \bbU,~w=U$, to the above equations is
\be
\bbU = (\bbpi - i\k\bar{z})g,
\sii
U = c(\bbz,\bbpi,g,\p),
\eq{CPLX_sol_chrEq}
\ee
where $c$ is an arbitrary function of the canonical variables.
The primary constraint is
\be
\vf \deff \p = 0,
\ee
and its c.c., which correspond to the lagrangian relation, $\del L/\del w = \del L/\del \bar{w}=0$.

The hamiltonian is
\be
H
&=&
\bbpi\cdot\bbU + \pi U
- 
\left(
\fr{\bbU^2}{2g} + i\k \bbU\cdot\bar{\bbz}
\right)
+
{\rm c.c.}
\nn
&=&
g\c + c\vf + {\rm c.c.},
\ee
where
\be
\c \deff \fr12(\bbpi - i\k\bar{\bbz})^2
=
\fr{\bbU^2}{2g}.
\ee
The equation of motion for a canonical variable, $q$,
is the canonical equation expressed as
\be
\dot{q} = \{q,H\},
\ee
where the Poisson bracket is defined by
\be
\{A,B\}
=
\fr{\del A}{\del \bbz}
\cdot
\fr{\del B}{\del \bbpi}
-
\fr{\del A}{\del \bbpi}
\cdot
\fr{\del B}{\del \bbz}
+
\fr{\del A}{\del g}
\fr{\del B}{\del p}
-
\fr{\del A}{\del p}
\fr{\del B}{\del g}.
\ee	
We denote
\be
A^{\sim} \deff \{A,H\},
\ee
which is frequently used in what follows.
For example the canonical equation is written as 
$\dot{q}=q^{\sim}$.

In order to compare the above relations to those of
the lagrangian theory, we define the pullback of
canonical variables by
\be
q_{\rm pb}
\deff
q|_{\bbpi = \bbW, ~\p = W},
\sii
\bbW \deff \fr{\del L}{\del \bbw},
\sii
W \deff \fr{\del L}{\del w} = 0.
\ee
The pullback of the canonical equations for $\bbpi$ and $\bbz$ are
\be
0
&=&
(\dot{\bbpi} - \bbpi^{\sim})_{\rm pb}
=
[{\bf EL}]_{u=U_{\rm pb}},
\\
0
&=&
(\dot{\bbz} - \bbz^{\sim})_{\rm pb}
=
\dot{\bbz} - \bbU_{\rm pb},
\ee
which are equivalent to the EL-equations, 
since $\bbU_{\rm pb}$
is dummy variable in this expression.\vss

The pullback of the primary constraint, $\vf=0$,
is identity, and it should be preserved in the orbit,
where EL-equations are satisfied.
The preservation of the primary constraint requires the 1st order secondary constraint which is expressed in the present case as
\be
\c = 0,
\eq{CPLX_secondC}
\ee
and its c.c..
Furthermore there is the 2nd order secondary constraint\footnote[2]
{
Here we call all constraints apart form the primary one as secondary constraints.
In a literature the secondary constraints of higher  order may be called, {\it e.g.}, tertiary constraints. 
}:
\be
\c_0 \deff
(\bbpi - i\k\bar{\bbz})
\cdot
(\bar{\bbpi} + i\k\bbz)
=
0,
\eq{CPLX_tertCon}
\ee
which preserves $\c = 0$.
There is no other constraints.

It has been proven, in general gauge theories, that 	 
the pullback of the secondary constraints to the lagrangian theory are all lagrangian constraints
\cite{hori_1812}.
In the present model we have the lagrangian constraints, (\ref{CPLX_LagC}) and (\ref{CPLX_2ndLC}), which correspond to the secondary constraints, (\ref{CPLX_secondC}) and (\ref{CPLX_tertCon}), respectively.
The 2nd order secondary constraint, (\ref{CPLX_tertCon}), plays an important role in deriving the critical dimension of the complex particle as is shown later.\vss

The primary and the secondary constraints in the present model  
are 1st-class, {\it i.e.}, their Poisson brackets 
are closed in themselves:
\be
&&
\{\vf,\c\}=\{\vf,\bar{\c}\} = 0,
\\
&&
\{\c,\bar{\c}\}=-2i\k\c_0,
\sii
\{\c,\c_0\} = -4i\k\c.
\eq{CPLX_PB_2ndC}
\ee
If we use the following basis
\be
L_1 = \fr{1}{2\k}\c,
\sii
L_{-1} = \fr{1}{2\k}\bar{\c},
\sii
L_0 = -\fr{1}{4\k}\c_0,
\ee
(\ref{CPLX_PB_2ndC})
are written in a compact form:
\be
\{L_n,L_m\} = i(n - m)L_{n+m},
\sii
(n,m = 0, \pm 1)
\ee
which makes apparent the sl$(2,\real)$ algebra 
of the constraint functions.\vss

Any function, $Q(q)$, of the canonical variables 
generates transformation through the Poisson bracket as $\d q = \{q,Q\}$, which in fact is one of the canonical transformations defined by Goldstein \cite{goldstein}. 
In particular if it satisfies 
\be
Q^{\sim} \equiv 0 
\sii {\rm mod}~\vf,
\eq{CPLX_Qtil=0}
\ee
it is called the canonical gauge transformation.

Consider the generating function, $Q$, 
which is a linear combination of the constraint
functions:
\be
Q = g\e\c + \bar{g}\bar{\e}\bar{\c} +  \e_0\c_0 +
\x\vf + \bar{\x}\bar{\vf},
\ee
where $\e,\e_0$ and $\x$ are infinitesimal parameters, 
then the transformation generated by $Q$ is
\be
\d_Q \bbz
&=&
g\e(\bbpi - i\k\bar{\bbz}) + \e_0(\bar{\bbpi} + i\k\b	bz),
\\
\d_Q g 
&=&
\x,
\ee
here we omit terms which vanish by the constraints.
The pullback of the above transformation to the
lagrangian theory is
\be
\d_{\rm L} \bbz
&=&
\e\bbU_{\rm pb} + \fr{\e_0}{\bar{g}}\bar{\bbU}_{\rm pb},
\\
\d_{\rm L} g
&=&
\x,
\ee
which is equivalent to the semi-gauge transformation, (\ref{CPLX_d_sg}).
We see by a simple calculation that the condition for the canonical geuge transformation, (\ref{CPLX_Qtil=0}), becomes
\be
&&
\e^{\sim}_0 - 2i\k g\bar{g}(\e - \bar{\e})
=
0,
\\
&&~~
\x
=
(g\e)^{\sim} + 4i\k\e_0 g,
\ee
the pullback of which precisely coincide with 
(\ref{CPLX_dot_e_0=}) 
and
(\ref{CPLX_gzi=}), 
the condition for the gauge transformation.

 \vss
From Eqs.(\ref{CPLX_consv_p}) and (\ref{CPLX_consv_p-pi}), we see the conserved momentum of the complex particle 
is expressed in terms of the canonical variables  
by
\be
\bbp = \bbpi + i\k\bar{\bbz}.
\ee
In closing this section we point out that
the Poisson bracket between the conserved 
momenta and its c.c. is
\be
\{(\bbp)^\m,(\bar{\bbp})^\n\}
=
2i\k \h^{\m\n},
\ee
which does {\it not} vanish for $\k \ne 0$.
This means that $\bbz$ and $\bar{\bbz}$ 
do not represent two independent quantities.

\section{Quantization}\eq{Quant}

In order to quantize the complex particle we represent the canonical variables as linear operator on some Hilbert space of state vectors which are expressed by functions of $\bbz$ and $\bar{\bbz}$. We assume the functions are single valued and bounded everywhere.
The inner product of two states $\f_1$ and $\f_2$ 
is defined by
\be
\bra\f_1|\f_2\ket = \int d^Dz d^D\bar{z}
~\f_1^*(z,\bar{z})\f_2(z,\bar{z}).
\ee
The operators corresponding to  canonical variables, $\bbpi,\bar{\bbpi}$, are
\be
\hat{\bbpi} 
= -i\bbdel \deff -i\pdel{}{\bbz} ,
\sii
\hat{\bar{\bbpi}} 
= -i\bar{\bbdel} \deff -i\pdel{}{\bar{\bbz}} .
\ee
The primary constraint function, $\vf=\p$, 
may be replaced by $-i\del/\del g$, 
but we set $\vf = 0$ simply by assuming the state function is independent of $g$.

The constraint functions, $L_n,~(n=0,\pm 1)$,
are replaced by the operators
\be
\hat{L}_1 &=&
\fr{1}{4\k}(-i\bbdel - i\k\bar{\bbz})^2,
\\
\hat{L}_{-1} &=&
\fr{1}{4\k}(-i\bar{\bbdel} + i\k \bbz)^2,
\\
\hat{L}_0 &=&
-\fr{1}{4\k}(-i\bbdel - i\k\bar{\bbz})\cdot(-i\bar{\bbdel} + i\k \bbz)
+
\a,
\ee
where 
$\a$ is a constant representing the ordering ambiguity of operators.
The commutation relation among $\hat{L}_n$ is
\be
[\hat{L}_n,~\hat{L}_m]
=
(n - m)
\left(
\hat{L}_{n+m}
-
\left(
\a - \fr{D}{4}
\right)
\d_{n+m}
\right).
\sii
(n,m = 0,\pm 1)
\eq{Quant_LL=}
\ee
According to the usual quantization scheme we
introduce the ghost variables, $c_n,~(n=0,\pm 1)$, 
and the BRST operator expressed as
\be
Q_{\rm BRST} =
\sum_{n=0,\pm 1}
c_n\hat{L}_n
-
\fr12
\sum_{n,m=0,\pm 1}
(n - m)c_n c_m\pdel{}{c_{n + m}} .
\ee
Requiring the nilpotency of $Q_{\rm BRST}$ we fix the ordering ambiguity as
\be
\a = \fr{D}{4}.
\ee
Then $\hat{L}_n$ becomes a basis of sl$(2,\real)$ 
without central term.\vss

In the classical theory the constraints assure the equivalence between the lagrangian and the canonical theories, and  are employed to define the physical subspace of whole phase space. In quantum theory they cannot be put as operator equations.
Then this condition is so relaxed that the matrix elements of any products of $\hat{L}_n$ between the physical states, 
$|{\rm phys}\ket = |\f\ket,|\vf\ket$, vanish:
\be
\bra\f|
\hat{L}_{n_1}\cdots\hat{L}_{n_k}
|\vf\ket = 0.
\ee
This condition is realized by the requirement
\be
\hat{L}_1|{\rm phys}\ket=\hat{L}_0|{\rm phys}\ket=0,
\eq{Qunnti_Lphys=0}
\ee
since 
 $\hat{L}_n,~(n=0,\pm 1)$ close under the 
commutation relations, and the fact, $\hat{L}^\dagger_1 = \hat{L}_{-1}$, 
giving 
$\bra{\rm phys}|\hat{L}_{-1}=0$.

In the quantum theory the classical conserved momentum  
$\bbp = \bbpi + i\k\bar{\bbz}$, 
is replaced by
\be
\hat{\bbp}
&=&
-i\bbdel + i\k\bar{\bbz},
\ee
and its hermitian conjugate.
In order that the model is physically meaningful 
there must be eigenvector of the conserved momentum with real eigenvalues in the physical sector of states.
Such a operator is 
$\hat{\bbP} = \b\hat{\bbp} + \bar{\b}\hat{\bbp^{\dagger}}$, 
with some complex value $\b$.
Since we can set $\b=1$ by a global rotation, 
we choose 
\be
\hat{\bbP} = \hat{\bbp} + \hat{\bbp}^\dagger,
\ee
as the conserved momentum without loss of generality.

Now let us obtain the function representing the physical state, which satisfies the constraint equations and
is the eigenvector of $\hat{\bbP}$.
For this purpose we must solve the following equations:
\be
&&
\hat{L}_1|k\ket = \hat{L}_0|k\ket = 0,
\\
&&
\hat{\bbP}|k\ket = \bbk|k\ket.
\eq{Quanti_Pk=kk}
\ee
Let us put
\be
|k\ket
=
e^{-\k\bbz\cdot\bar{\bbz}}f(z,\bar{z}).
\ee
Then the condition $\hat{L}_1|k\ket=0$ becomes $\bbdel^2f(z,\bar{z}) = 0$.
The eigenvalue equation 
(\ref{Quanti_Pk=kk})
is written in terms of $f(z,\bar{z})$ as
\be
(\bbdel + \bar{\bbdel} - 2\k\bar{\bbz} - i\bbk)f(z,\bar{z}) = 0.
\eq{CPLX_()f=0}
\ee
This is the form of separating variables with respect to $\bbz$ and $\bar{\bbz}$.
The solution is written as
\be
f(z,\bar{z})
=
e^{i\bbk_1\cdot\bbz 
+ 
i\left(\bbk - \bbk_1\right)\cdot\bar{\bbz} + \k\bar{\bbz}^2}
\ee
where $\bbk_1$ is an arbitrary separation constant.
The general solution is obtained by multiplying 
arbitrary function, $a(k_1)$, to $f(z,\bar{z})$ 
and integrating over $k_1$:
\be
f(z,\bar{z})
=
e^{i\bbk\cdot\bar{\bbz} + \k\bar{\bbz}^2}g(y),
\sii
\bby = i(\bbz - \bar{\bbz}),
\eq{Quanti_genSol1}
\ee
where $g(y)$ is the function of real variable $\bby$,
 whose Fourier coefficient is $a(k_1)$.
Since $\bbdel^2f(z,\bar{z}) = 0$,
$g(y)$ satisfies 
\be
\fr{\del^2}{\del y_\m\del y^\m} g(y) = 0.
\eq{deldelg=0}
\ee

At last the condition 
$\hat{L}_0|k\ket=0$
is expressed as
\be
(\bbdel\cdot\bar{\bbdel} - 2\k\bbz\cdot\bbdel - 4k\a)
f(z,\bar{z}) =0.
\ee
Substituting 
(\ref{Quanti_genSol1})
into this relation,
we get
\be
\left[
\left(y^\m - \fr{k^\m}{2\k}\right)\pdel{}{y^\m} 
\right]
g(y)
=
0.
\eq{Quanti_(cdots)g=0}
\ee
Defining function, $h(u)$, by
\be
g(y)
=
\left[
\left(
\bby - \fr{\bbk}{2\k}
\right)^2
\right]^{-\a}
h
\left(
y - \fr{k}{2\k}
\right),
\ee
we see that Eq.(\ref{Quanti_(cdots)g=0}) 
is written as
\be
u^\m\pdel{}{u^\m} h(u) &=& 0.
\eq{Quanti_uDuh=0}
\ee
Using this equation and (\ref{deldelg=0}), we see 
\be
\left(\dal_u - \fr{K}{\bbu^2}\right)h(u) = 0,
\eq{Quanti_(dal - k/u2)h=0}
\sii
K = 2\a(D-2(\a + 1)) = \fr14 D(D-4),
\eq{Quanti_K=D(D-4)/4}
\ee
where
$\dal_u = \del^2/\del u^\m\del u_\m$ is 
the d'Alembert operator  with respect 
to $D$-dimensional real coordinates, $u^\m$. \vss

Our task is to solve Eqs.
(\ref{Quanti_uDuh=0})
and
(\ref{Quanti_(dal - k/u2)h=0})
with some physical assumptions.
In order to see the physical meaning of these  equations, let us remember the corresponding equations in the case of single relativistic particle.
In that case there is one secondary constraint, $\bbpi^2=0$, and the corresponding quantum equation  is 
$\dal|k\ket=0$. The solution is the plane wave, $|k\ket = e^{i\bbk\cdot\bbx}$, with $\bbk^2=0$.
In the case of the complex particle the corresponding state function is
\be
|k\ket
&=&
e^{i(\bbk + \k\bby )\cdot\bar{\bbz}}
\left[
\left(
\bby - \fr{\bbk}{2\k}
\right)^2
\right]^{-D/4}
h
\left(
\bby - \fr{\bbk}{2\k}
\right),
\sii
y=i(\bbz - \bar{\bbz}),
\ee
as is seen by the definition of $h$.
A natural assumption for the function $h(u)$ 
is that it is  single-valued and bounded everywhere.

The solution for $h(u)$ is obtained by using 
the eigenvalue equation of the Laplace-Beltrami(LB) operator on $S^{1,D-2}$ and $S^{D-1}$ in the case of Minkowski and Euclid metric, respectively.
LB is connected with the (pseudo-)rotation in the real spacetime, which is generated by
\be
\ell^{\m\n}
=
-i\left(
u^{\m}\pdel{}{u_\n} - u^{\n}\pdel{}{u_\m}
\right),
\ee
acting on the real coordinates $u^\m$ as
\be
\left(\fr{i}{2}\e_{\m\n}\ell^{\m\n}\right) u_{\l}
=
\e_{\m\l}u^{\m},
\sii \e_{\n\m} = - \e_{\m\n},
\ee
where  $\e_{\n\m}$ are infinitesimal anti-symmetric parameters.
$\ell^{\m\n}$ satisfy the commutation relation
\be
[\ell^{\m\n},\ell^{\m'\n'}]
=
i
(\h^{\n'[\m}\ell^{\n]\m'} - \h^{\m'[\m}\ell^{\n]\n'}).
\ee
The definition of LB is\footnote[2]{
In the case of $D=3$ Euclidean space 
${\vDal}_{\rm LB}$ is minus square of the angular momentum operators. Here the minus sign is put in accordance with mathematical literatures.
}
\be
{\vDal}_{\rm LB} 
\deff
-\fr12\ell_{\m\n}\ell^{\m\n},
\ee
and it commutes with all $\ell^{\m\n}$,
{\it i.e.}, one of Casimir operators of the (pseudo-)rotation group.

LB has properties which are easily treated in the 
(pseudo-)spherical coordinates, $(r,\th_1,..,\th_{D-1})$, as described in the next section.
An important relation between the d'Alembert operator  and 
LB in the (pseudo-)spherical coordinates is
\be
s~\dal_u 
=
\fr{1}{r^2}{\vDal}_{\rm LB} 
+ \fr{1}{r^{D-1}}\pdel{}{r} r^{D-1}\pdel{}{r} , 
\eq{Quanti-dal-vDal}
\ee
where
\be
s \deff {\rm sgn}(u_{\m}u^\m).
\ee
The proof of (\ref{Quanti-dal-vDal}) 
is simple, but we give it
in {\bf Appendix \ref{App_xx_dal-vDal}}, 
for completeness.
The condition 
(\ref{Quanti_uDuh=0})
becomes $r$ independence of $h(u)$ in the 
(pseudo-)spherical coordinates
(see (\ref{LB_f6}) in the next section).
Hence the second term in r.h.s. of 
(\ref{Quanti-dal-vDal})
vanishes in acting to $h(u)$.
In Section 6 we solve the eigenvalue equation of ${\vDal}_{\rm LB}$,
\be
{\vDal}_{\rm LB} h = \l h,
\ee
where the eigenvalue $\l$ must have restricted values. 
Therefore, from Eq.
(\ref{Quanti_(dal - k/u2)h=0}) we get the relation between  the dimension of spacetime and the eigenvalue $\l$,
\be
\fr{1}{4}D(D-4) = \l,
\eq{Quanti_D=l}
\ee
here we use $\bbu^2 = sr^2$.

  \vss

\section{Laplace-Beltrami operator}

In the theory of spherical harmonics the LB operator plays an important role, which maps $S^{D-1}$ 
into itself in the case of the Euclid metric.
For the purpose of the present paper it is 
convenient to express LB in terms of the 
(pseudo-)spherical coordinates. \vss


The relation between the $D$-dimensional orthogonal coordinates, $u^{\m}$, in the Euclid spacetime, and the 
spherical coordinates, $(r,\th_1,..,\th_{D-1})$, is defined by
\be
u^m &=& rC_{m + 1}P_m,
\sii (0 \le m \le D-2),
\eq{LB_SCoord}
\\
u^{D-1} &=& rP_{D-1},
\ee
here we use the abbreviated notation
\be
C_m \!\! &\deff& \!\!\cos{\th_m},
\sii
S_m \deff \sin{\th_m},
\\
P_m \!\! &\deff& \!\! \prod_{k=1}^mS_k,
\si
(1\le m \le D-1),
\sii
P_0 \deff 1.
\ee
It is convenient to set $C_D=1,~S_D=0$, since then only (\ref{LB_SCoord}) is necessary in that notation.
The ranges of the variables in the spherical coordinates are
\be
0 \le r < \infty,
\sii
0 \le \th_m \le \p,
\si (1 \le m \le D-2),
\sii
0 \le \th_{D-1} \le 2\p,
\ee
here we must set $0 \le \th_1 \le 2\p$ in the case of $D=2$.

The pseudo-spherical coordinates in the Minkowski  spacetime can be defined in a similar manner.
First we divide the whole spacetime into three connected subspaces:
\be
M_{+} &=& \{u|~ u_\m u^\m \le 0,~u^0 \ge 0\},
\eq{LB_defM+}
\\
M_{-} &=& \{u|~ u_\m u^\m \le 0,~u^0 \le 0\},
\eq{LB_defM-}
\\
M_0   &=& \{u|~ u_\m u^\m \ge 0\}.
\eq{LB_defM0}
\ee
The set of points whose radius vector from the origin is time-like belong to $M_{\pm}$, and 
those of space-like ones belong to $M_0$.
The boundaries are  the upper light-cone for $M_+$, 
the lower light-cone for $M_-$ and  both of them for $M_0$.
For each connected subspace the pseudo-spherical coordinates are defined by
\be
u^m = rC_{m + 1}P_m,
\si
(0 \le m \le D-1),
\eq{LB_PSCoordm_Mpm}
\ee
where 
\be
&&
(C_1, S_1) = 
\kakkob (\pm\cosh{\t},~\pm\sinh{\t}) \sii {\rm for}~ u\in M_{\pm}\\
   (\sinh{\t}, ~-\cosh{\t}) \sii\si {\rm for}~ u\in M_{0}
\kakkoe
, 
\sii
\t = \th_1,
\eq{LB_PScoord_C_1_S_1}
\\
&&
C_k \deff \cos{\th_k},
\sii
S_k \deff \sin{\th_k},
\sii
(2\le k \le D-1)
\\
&&
P_m \deff \prod_{k=1}^{m} S_k,
\sii
(1 \le m \le D-1),
\\
&&\sii
C_{D} = 1,
\sii  S_{D} = 0,
\sii P_0 = 1.
\ee
Here we use $\t$ instead of $\th_1$, 
since $\t$ is the argument of hyperbolic functions.
The ranges of $(r,\t.\th_k),~(2\le k \le D-1)$ 
are
\be 
 0 < r < \infty,
\si
-\infty < \t < \infty,
\si
 0\le \th_k \le \p,
\si
(2\le k \le D-2),
\si
 0\le \th_{D-1} \le 2\p.
\eq{LB_PSCoord_range}
\ee

Let us list up some useful formulas with simple proofs of them.
\be
\bullet  &&\si
\h_{00}C_1^2 + S_1^2 = s,
\sii
\del_1\vecb C_1 \\
            S_1
      \vece 
=
      s\vecb -S_1 \\
          \h_{00}C_1
      \vece ,
\sii
\del_m \deff \pdel{}{\th_m} ,
\eq{LB_f1}
\\
\bullet  &&\si
\sum_{k=m}^{D-1}C_{k+1}^2P_k^2 = P_m^2,
\sii
(1 \le m \le D-1)
\eq{LB_f2}
\\
\bullet  &&\si
r = \tilde{r} 
~\deff~ 
\sqrt{s~u_\m u^\m}~,
\sii
s \deff {\rm sgn}(u_\m u^\m),
\eq{LB_f3}
\\
\bullet  &&\si
C_{1}
=
\fr{u^{0}}{\sqrt{s~u_\m u^\m}}~,
\sii
\eq{LB_C1=u^0}
\eq{LB_f4}
\\ 
\bullet  &&\si 
C_{n}
=
\fr{s_{n-1}u^{n-1}}{
\sqrt{\sum_{k=n-1}^{D-1}u_k^2}}~,
\sii
(2 \le n \le D-1),
\sii
s_n \deff {\rm sgn}(P_n).
\eq{LB_f5}
\ee
 \noi\sen

\noi pr.)

Eqs.(\ref{LB_f1})
are apparent from the definitions.
Eq.(\ref{LB_f2})
is derived by summing the square of
 $C_{k+1}P_k,~(k\ge 1)$
from  $k=D-1$ to  $k=m,~(m\ge 1)$.
Using this relation with $m=1$
we have
\be
s~u_\m u^\m = s~\left(u_0u^0 + 
\sum_{k=1}^{D-1}(u_k)^2\right)
=
sr^2(\h_{00} C_1^2 + S_1^2)
=
r^2,
\eq{LB_pr_f3}
\ee
hence  
(\ref{LB_f3})
follows.
From  
(\ref{LB_f1})
and
$u^0 = rC_1$ 
we have
(\ref{LB_f4}).
Finally, for 
$n \ge 2$
we have
\be
C_n 
= 
\fr{u^{n-1}}{rP_{n-1}}
=
\fr{s_{n-1}u^{n-1}}{r|P_{n-1}|},
\ee
and this coincides with
(\ref{LB_f5})
by
(\ref{LB_f2}). \qed\\
\noi\sen

 \vss
 
The derivative operators of $u^\m$ are
expressed by those of $(r,\th_1,..,\th_{D-1})$ as
\be
\bullet  &&\si
\fr{u^\m}{\tilde{r}}\pdel{}{u^\m} = \pdel{}{r},
\eq{LB_f6}
\\
\bullet  &&\si
\pdel{}{u_0}
\! = \!
sC_{1}\pdel{}{r}
-
\h_{00}\fr{S_1}{r}\del_1
,
\sii
\del_m \deff \pdel{}{\th_m} ,
\eq{LB_f7}
\\
\bullet  &&\si
\pdel{}{u_{m}}
\! = \!
C_{m+1}P_{m}
\left(
s\pdel{}{r} 
+ 
\fr{1}{r}\sum_{p=1}^{m}\fr{C_pS_p}{P_{p}^2}\del_{p}
\right)
-
\fr{S_{m+1}}{rP_m}\del_{m+1},
\si (1 \le m \le D-1)
\eq{LB_f8}
\ee
where
\be
&&
C_D \! = \! 1,\sii S_D = 0,
\sii 
P_m \deff \prod_{k=1}^m S_k,
\sii
P_0 = 1.
\eq{LB_f6-8_def}
\ee

\noi\sen

\noi pr.)

Eq.(\ref{LB_f6}) is apparent from the definition.
As is seen from 
(\ref{LB_f5})
$C_p$ does not depend on $u^0,..,u^{p-2}$.
Hence we have 
for
$1 \le m \le D-2$
\be
\pdel{}{u_{m}}
&=&
\fr{\del \tilde{r}}{\del u_{m}}\pdel{}{r}
+
\left(\pdel{}{u_m} \fr{u^0}{\sqrt{su_\m u^\m}}\right)
\left(\fr{d C_{1}}{d \th_1}\right)^{-1}\del_{1}
+
\sum_{p=2}^{m+1}
\left(
\pdel{}{u_{m}}
\fr{s_{p-1}u^{p-1}}{\sqrt{\sum_{k=p-1}^{D-1}u_ku^k}}
\right)
\left(\fr{d C_{p}}{d \th_p}\right)^{-1}\del_{p}.
\nn
\eq{LB_pr_f6-8}
\ee
The second term in r.h.s. of this equation is
\be
({\rm 2nd}~{\rm term}~{\rm in}~{\rm r.h.s.})
&=&
\fr{u^0u^m}{r^3S_1}\del_1.
\ee
By a straightforward but somehow lengthy calculation, using the above equation, 
we have (\ref{LB_f8}) in the case of  $(1 \le m \le D-2)$.

In the case of $m=0$ we have
\be
\pdel{}{u_0}  
=
\fr{\del \tilde{r}}{\del u_0}\pdel{}{r}
+
\left(\pdel{}{u_0} \fr{u^0}{\sqrt{su_\m u^\m}}\right)
\left(\fr{d C_{1}}{d \th_1}\right)^{-1}\del_{1},
\ee
here the 2nd term of r.h.s. is
\be
({\rm 2nd}~{\rm term}~{\rm in}~{\rm r.h.s.})
&=&
\fr{\h_{00}s}{\tilde{r}^3}
(u_\m u^\m - u_0u^0)
\left(\fr{d C_{1}}{d \th_1}\right)^{-1}\del_{1}
\nn
&=&
-
\fr{\h_{00}}{\tilde{r}}
\left(
\sum_{m=1}^{D-1}C_{m+1}^2P_m^2
\right)
\left(S_1\right)^{-1}\del_{1}.
\ee
Using 
(\ref{LB_f2})
we see
\be
\si
=
-\h_{00}\fr{S_1}{r}\del_1,
\ee
hence
(\ref{LB_f7}) follows.

The case of 
$m=D-1$ 
is the same as that of
$m\le D-2$
apart from that
the upper limit of the sum in
(\ref{LB_pr_f6-8})
is $m$.
Hence we have
\be
\pdel{}{u_{D-1}}
&=&
P_{D-1}
\left(
s\pdel{}{r} 
+ 
\fr{1}{r}
\sum_{p=1}^{D-1}\fr{C_pS_p}{P_p^2}\del_p
\right).
\ee
After all (\ref{LB_f8}) holds for 
$(0 \le m \le D-1)$. \qed\\
\noi\sen

 \vss

Since, in the pseudo-spherical coordinates,  
$C_1,~S_1$
are the hyperbolic function,
they may appear in various relations in different way from 
$C_p,~S_p,~(p\ne 1)$.
For example 
Eq.(\ref{LB_f8})
 with $m=0$
does not coincide with
(\ref{LB_f7}).\vss

The main purpose of this section is to
represent the LB operator in the (pseudo-)spherical coordinates.
In the case of Euclid spacetime the expression of 
the LB operator, and the proof of it in another context, have been given in 
\cite{dai}-\cite{nomura}. Here we extend them to the case of Minkowski spacetime.
The result is
\be
{\vDal}_{\rm LB}
=
\sum_{n=1}^{D-1}
\fr{\s_n S_n^{n - D +3}}{P_n^2}
\del_n S_n^{D - n - 1}\del_n,
\sii
\s_n \deff \kakkob \h_{00}\sii {\rm for}~n=1 \\
               ~ s     \sii~ {\rm for}~n \ge 2
       \kakkoe .
\eq{LB_LB_in_PSCoord}
\ee
If we put $\s_n=1$ in this equation we get  ${\vDal}_{\rm LB}$ 
in the Euclid metric.
Using 
(\ref{LB_f8})
and
the definition of $\ell^{\m\n}$, 
it might be possible to obtain
(\ref{LB_LB_in_PSCoord}).
Since the direct substitution of $\ell^{\m\n}$
into the definition of ${\vDal}_{\rm LB}$ 
gives very tedious expression to handle,
it may not be a smart method to prove (\ref{LB_LB_in_PSCoord}) (if not impossible). 
We give here a proof by the induction with respect to the dimension $D$.

In the case of $D=2$ we see
\be
&&
u^0 = rC_1, \si u^1 = rS_1, 
\si 
\pdel{}{u_0}
= C_1\pdel{}{r} - \fr{\h_{00}S_1}{r}\del_1, 
\si 
\pdel{}{u_1}
= S_1\pdel{}{r} + \fr{C_1}{r}\del_1,	 
\\
&&
\ell^{01} 
=
-ir
\left[C_1 
\left( S_1\pdel{}{r} + \fr{C_1}{r}\del_1\right)
-
S_1
\left(C_1\pdel{}{r} - \fr{\h_{00}S_1}{r}\del_1\right)
\right]
\nn
&& \si~\!
=
-i(C_1^2 + \h_{00}S_1^2)\del_1
=
-is\h_{00}\del_1.
\ee
Hence we have 
${\vDal}_{\rm LB}= -\ell^{01}\ell_{01}= \h_{00}\del_1^2$, 
and Eq.(\ref{LB_LB_in_PSCoord})
holds.
Let us prove 
Eq.(\ref{LB_LB_in_PSCoord})
in $D$-dimension by assuming it holds 
in $(D-1)$-dimension.

For that purpose it is convenient to 
define the (pseudo-)spherical coordinates through the
following two steps:
\be
u^m = \r C_{m+1}P_m,
\sii
(1 \le m \le D-1),
\eq{LB_PScord2-1}
\ee
and
\be
u^0 &=& rC_1,
\sii\sii\sii\sii\sii\sii\si
\eq{LB_PScord2-21}
\\
\r &=& rS_1,
\eq{LB_PScord2-22}
\ee
where $C_1,~S_1$  are defined by
(\ref{LB_PScoord_C_1_S_1}).
The two Eqs.
(\ref{LB_PScord2-21})
and
(\ref{LB_PScord2-22})
are the same as the relation between two-dimensional 
orthogonal coordinates $(u^0,\r)$ and the (pseudo-)spherical coordinates $(r,\th_1)$.
Since in the case of the Minkowski metric,
$C_1,~S_1$ are hyperbolic functions,
and being different from the spacial part of
 (\ref{LB_PScord2-1}), 
 this two step definition makes the calculations easy.

Let us write again the relation 
(\ref{Quanti-dal-vDal}),
\be
s~\dal^{(D)}
=
\fr{{\vDal}_{\rm LB}^{(D)}}{r^2}
+
\fr{1}{r^{D-1}}\pdel{}{r} r^{D-1} \pdel{}{r} ,
\eq{LB_dal-vDal_again}
\ee
which holds in arbitrary dimension, $D$.
For the spacial part of the 
(pseudo-)spherical coordinates 
in the definition of the above two step 
definition, 
which is $(D-1)$-dimensional Euclid space ($s=+1$), 
Eq.(\ref{LB_dal-vDal_again})
is written as
\be
\dal^{(D-1)}
=
\fr{{\vDal}_{\rm LB}^{(D-1)}}{\r^2}
+
\fr{1}{\r^{D-2}}\pdel{}{\r} \r^{D-2} \pdel{}{\r} .
\eq{LB_dal-vDal_D-1}
\ee
For the two dimensional part of the (pseudo-)spherical coordinates, 
it is written as
\be
s
\left(
\h_{00}
\fr{\del^2}{\del u_0^2}
+
\fr{\del^2}{\del \r^2}
\right)
=
\fr{\h_{00}}{r^2}\del_1^2 
+ 
\fr{1}{r}\pdel{}{r} r\pdel{}{r} .
\eq{LB_dal-vDal_D=2}
\ee

From Eqs.
(\ref{LB_dal-vDal_again})
$-$
(\ref{LB_dal-vDal_D=2}), 
and $\dal^{D} = \dal^{(D-1)} + \del^2/\del \r^2$,
we have
\be
{\vDal}_{\rm LB}^{(D)}
&=&
sr^2
\left(
\fr{{\vDal}_{\rm LB}^{(D-1)}}{\r^2}
-
\fr{s}{r^{D-1}}\pdel{}{r} r^{D-1} \pdel{}{r} 
+
\fr{1}{\r^{D-2}}\pdel{}{\r} \r^{D-2} \pdel{}{\r} 
+
\fr{s}{r}\pdel{}{r} r\pdel{}{r} 
+
\fr{s\h_{00}}{r^2}\del_1^2
-
\fr{\del^2}{\del\r^2}
\right)
\nn
&=&
sr^2
\left(
\fr{{\vDal}_{\rm LB}^{(D-1)}}{\r^2}
-
\fr{s(D-1)}{r}\pdel{}{r} 
	+
\fr{D-2}{\r}\pdel{}{\r} 
+
\fr{s\h_{00}}{r^2}\del_1^2
\right).
\eq{LB_vDal=1}
\ee
From the definition of the two dimensional part of 
the (pseudo-)spherical coordinates, we see $\r=rS_1$ 
and the differential operator with respect to $\r$
is obtained by putting $\r=u^1,~C_2=1,~S_2=0$
in Eq.(\ref{LB_f8}),
\be
\pdel{}{\r}
=
sS_1\pdel{}{r} + \fr{C_1}{r}\del_1.
\ee
Substituting it to (\ref{LB_vDal=1}), 
we see the dependence on $r$ and the terms of 
$\del/\del r$ cancels out, 
giving 
\be
{\vDal}_{\rm LB}^{(D)}
&=&
\fr{s}{S_1^2}
{\vDal}_{\rm LB}^{(D-1)}
+
\fr{s(D-2)C_1}{S_1}\del_1
+
\h_{00}\del_1^2
\nn
&=&
\fr{s}{S_1^2}
{\vDal}_{\rm LB}^{(D-1)}
+
\fr{\h_{00}}{S_1^{D-2}}\del_1 S_1^{D-2}\del_1.
\eq{LB_vDal=2}
\ee
By the assumption of the deduction 
Eq.(\ref{LB_LB_in_PSCoord}) holds
in $(D-1)$-dimensional Euclid spacetime,
hence we have
\be
{\vDal}_{\rm LB}^{(D-1)}
=
\sum_{n=1}^{D-2}
\fr{S_n^{n - D +4}}{P_n^2}
\del_n S_n^{D - n - 2}\del_n
=
S_1^2
\sum_{n=2}^{D-1}
\fr{S_n^{n - D + 3}}{P_n^2}
\del_n S_n^{D - n - 1}\del_n.
\ee
Substituting it to Eq.(\ref{LB_vDal=2})
we get
\be
{\vDal}_{\rm LB}^{(D)}
&=&
s
\sum_{n=2}^{D-1}
\fr{S_n^{n - D + 3}}{P_n^2}
\del_n S_n^{D - n - 1}\del_n
+
\fr{\h_{00}}{S_1^{D-2}}\del_1 S_1^{D-2}\del_1
\nn
&=&
\sum_{n=1}^{D-1}
\fr{\s_{n}S_n^{n - D + 3}}{P_n^2}
\del_n S_n^{D - n - 1}\del_n.
\ee
Hence 
(\ref{LB_LB_in_PSCoord}) holds in $D-$dimensional Minkowski as well as Euclid spacetime. \qed\\
\noi\sen

 \vss

The expression  (\ref{LB_LB_in_PSCoord}) 
for the Minkowski spacetime 
has not been published in a mathematical literature 
as far as we know.
In the next section we find all possible eigenvalues of ${\vDal}_{\rm LB}$, 
and deduce the critical dimension of spacetime.

\section{Critical dimension}

In a mathematical literature the eigenvalue problem of LB is often discussed through d'Alembert equation, $\dal \Ps = 0$.
In this section we 
solve the eigenvalue equation of ${\vDal}_{\rm LB}$ 
directly in terms of the angle variables in the (pseudo-)spherical coordinates, 
since ${\vDal}_{\rm LB}$ is expressed  only by those variables
irrelevant to  the radius, $r$.

Let us solve the eigenvalue equation of ${\vDal}_{\rm LB}$:
\be
{\vDal}_{\rm LB} \Ps = \l\Ps.
\eq{CD_eigenEq}
\ee
We assume the eigen-function $\Ps$ 
is single-valued and bounded everywhere.
The eigenvalue $\l$ is related to the dimension of 
spacetime $D$ as Eq.(\ref{Quanti_D=l}).

In the case of $D=2$ the eigenvalue equation is
\be
\h_{00}\del_1^2\Ps = \l\Ps,
\ee
and the solution is\footnote[2]{
Here and hereafter we omit the normalization  constant of $\Ps$.}
\be
\Ps = e^{m\th_1}, 
\sii \l = \h_{00}m^2.
\ee
For the Euclid metric ($\h_{00}=1$) the range of the spherical coordinate, $\th_1$, is $0 \le \th_1 \le 2\p$.
Hence, from the single-valuedness of $\Ps$, 
we see
\be
m = iL,
\sii
L \in \inte,
\sii
\l = -L^2 \le 0,
\sii	
({\rm for}~{\rm Euclid}~{\rm metric}).
\eq{CD_Euclid_l=}
\ee
On the other hand for the Minkowski metric ($\h_{00}=-1$) the 
range of the pseudo-spherical coordinate, $\t=\th_1$, 
is
$-\infty < \t < \infty$.
Hence, from the boundedness of $\Ps$ at infinity ($\t=\pm\infty$), we see 
$\Re m = 0$,
{\it i.e.},
\be
m = iL,
\sii
L \in \real,
\sii
\l = L^2 \ge 0.
\sii
({\rm for}~{\rm Minkowski}~{\rm metric}).
\eq{CD_Minkow_l=}
\ee
Thus the sign of the eigenvalue depends on 
whether the invariant space of (pseudo-)rotation 
is compact ($S^{D-1}$) or non-compact ($S^{1,D-2}$).

Next let us consider the case of $D\ge 3$.
The eigenvalue equation is
\be
\left(
\sum_{n=1}^{D-1}
\fr{\s_nS_n^{n - D +3}}{P_n^2}
\del_n S_n^{D - n - 1}\del_n
\right)
\Ps
=
\l\Ps,
\eq{CD_EVE_Dge3}
\ee
which is of the form of separated variables 
with respect to 
$(\th_1,..,\th_{D-2})$
and
$\th_{D-1}$.	
Putting 
$\Ps = G_1(\th_1,..,\th_{D-2})M_1(\th_{D-1})$, 
Eq.(\ref{CD_EVE_Dge3})
reduces to 
\be
\left(
\sum_{n=1}^{D-2}
\fr{\s_nS_n^{n - D + 3}}{P_n^2}
\del_n S_n^{D - 1 - n}\del_n
-
\l
+
\fr{c_1}{P_{D-2}^2}
\right)
G_{1} 
&=& 0,
\eq{CD_EVE4G_1}
\\
\left(
s\del_{D-1}^2 - c_1
\right)M_1 
&=&
0,
\eq{CD_EVE4M_1}
\ee
where $c_1$ is the separation constant.
If $D=3$ the solving step is completed here.
But for $D\ge 4$ Eq.(\ref{CD_EVE4G_1}) should be 
solved, which is again of the separated type with 
respect to 
$(\th_1,..,\th_{D-3})$
and
$\th_{D-2}$.
Putting 
$G_1 = G_2(\th_1,..,\th_{D-3})M_2(\th_{D-2})$,
we have
\be
\left(
\sum_{n=1}^{D-3}
\fr{\s_nS_n^{n - D + 3}}{P_n^2}
\del_n S_n^{D - 1 - n}\del_n
-
\l 
+
\fr{c_2}{P_{D-3}^2}
\right)
G_{2} 
&=& 0,
\eq{CD_EVE4G_2}
\\
\left(
\fr{s}{S_{D-2}}\del_{D-2}S_{D-2}\del_{D-2}
+
\fr{c_1}{S_{D-2}^2} - c_2
\right)M_2 
&=& 0,
\eq{CD_EVE4M_2}
\ee
where $c_2$ is the separation constant.

Going on similarly until $m=D-2$ we get the following 
set of ordinary differential equations:
\be
\left(
\h_{00}S_1^{2 - D}
\del_1 S_1^{D - 2}\del_1
-
\l 
+
\fr{c_{D-2}}{P_{1}^2}
\right)G_{D-2} &=& 0,
\eq{CD_EVE4G_m}
\\
\left(
\fr{s}{S_{D-m}^{m-1}}\del_{D-m}S_{D-m}^{m-1}\del_{D-m}
+
\fr{c_{m-1}}{S_{D-m}^2} - c_{m} 
\right)M_{m}
&=& 0,
\sii
(m = 3,..,D-2),
\eq{CD_EVE4M_m}
\ee
where 
$G_{m} = G_{m+1}(\th_1,..,\th_{D -m - 2})M_{m+1}(\th_{D - m-1}),~(m=3,4,..,D-3)$, 
and
$c_m,~(m=3,..,D-2)$
are the separation constants.
Since 
(\ref{CD_EVE4M_1})
and
(\ref{CD_EVE4M_2})
coincide with 
(\ref{CD_EVE4M_m})
if we put 
$c_0 = 0,~m=1,2$ there, 
the equations to be solved are
(\ref{CD_EVE4G_m})
and
(\ref{CD_EVE4M_m}) with $(m=1,2,..,D-2)$, 
the total number of which is $D-1$.
Using the above solutions, the eigen-function of 
${\vDal}_{\rm LB}$ is written as
\be
\Ps = 
G_{D-2}\prod_{m=1}^{D-2}M_m.
\eq{CD_EF_Ps=}
\ee

 \vss
Now let us solve 
(\ref{CD_EVE4G_m})
and
(\ref{CD_EVE4M_m}) with $(m=1,2,..,D-2)$.
As a solution to Eq.(\ref{CD_EVE4M_m}) for $D\ge 3$
put 
\be
M_m(\th_{D-m}) 
=
\kakkob e^{in_1\th_{D-1}}, \sii\sii (m=1) \sii~~ \\
 S_{D-m}^{n_m}(\th_{D-m}), \sii (2 \le m \le D-2) 
\kakkoe,
\eq{CD_sSol2_Mm}
\ee
where $n_m,(m=1,..,D-2)$ are complex numbers.
Then the equations reduce to
\be
&&
(sn_1^2 + c_1)e^{n_1\th_{D-1}} = 0,
\\
&&
(sn_m(n_m + m - 2) + c_{m-1})
S_{D-m}^{n_m-2}
-
(sn_m(n_m + m - 1) + c_m)
S_{D-m}^{n_m}
=
0.
\nn
&& 
\sii\sii\sii\sii
\sii\sii\sii\sii
\sii\sii
 (2 \le m \le D-2)
\ee
Hence if 
$n_m,~(1 \le m \le D-2)$
satisfy
\be
n_1^2 &=& -sc_1,
\eq{CD_MmRed2-1}
\\
c_{m} &=& -sn_{m+1}(n_{m+1} + m - 1),
\si
(1 \le m \le D-3)
\eq{CD_MmRed2-2}
\\
c_m &=& -sn_m(n_m + m - 1),
\sii\si
(2 \le m \le D-2),
\eq{CD_MmRed2-3}
\ee
for given 
$c_m,(1 \le m \le D-2)$, 
then
$M_m$ are solutions of
(\ref{CD_EVE4M_m}) with $(m=1,2,..,D-2)$.
Note that Eqs.
(\ref{CD_MmRed2-1})
$-$
(\ref{CD_MmRed2-3})
do not change by replacing $n_m$ to $-(n_m + m - 1)$ 
in them.
Thus other solutions are obtained by the 
above replacement.
Since Eq.(\ref{CD_EVE4M_m}) are homogeneous ordinary differential equations of second rank,
the dimension of the solution space as a vector space is two.
The above two special solutions are 
linearly independent, hence the general solutions are  
linear combinations of them.

Since $S_m(\th_m)=\sin{\th_m}$ for $m \ge 2$, 
the boundedness at origin requires
$\Re n_m \ge 0,~(2 \le m \le D-2)$.
Also the single-valuedness of the solution 
requires $n_1\in \inte$.
After all, $n_m$ which satisfy the boundedness 
and the single-valuedness are determined by
\be
&&
n_1^2 = n_2^2,
\sii
n_1\in \inte
\eq{CD_cond_for_n_1=n_2}
\\
&&
n_{m+1}(n_{m+1} + m - 1)
=
n_m(n_m + m - 1),
\sii
(2 \le m \le D-2),
\eq{CD_cond_for_n_m+1=n_m}
\\
&&
\sii
\Re n_m \ge 0,
\sii
(2 \le m \le D-2).
\eq{CD_cond_for_n_m_ge2}
\ee
Firstly from 
(\ref{CD_cond_for_n_1=n_2})
and 
$\Re n_2\ge 0$
we have
$n_2=|n_1| \in \inte_{\ge 0}$.
Since (\ref{CD_cond_for_n_m+1=n_m}) is written as
\be
(n_{m+1} - n_m)(n_{m+1} + n_m + m - 1) = 0, 
\ee
it has two solutions 
$n_{m+1}=n_m$ and $n_{m+1}=-n_{m} -m +1$
as a 2nd order algebraic equation for $n_{m+1}$.
For $m=2$ the latter solution
is excluded because 
$\Re n_3 < 0$.
Similarly we see the possible solutions are
\be
\ell \deff |n_1| = n_2 = \cdots = n_{D-2} \in \inte_{\ge 0}.
\eq{CD_n_1=...=n_D-2}
\ee

In order to solve the remaining Eq.(\ref{CD_EVE4G_m}) 
we put $G_{D-2}(\th_1) = S_1^p(\th_1)$ with 
a complex number $p$.
Substituting it to 
Eq.(\ref{CD_EVE4G_m}), 
we have
\be
[s~p(D-3+p) + c_{D-2}]S_1^{p-2}
-
[p(D-2+p) + \l]S_1^p
=
0.
\ee
Hence we see that
$G_{D-2}(\th_1) = S_1^p(\th_1)$
is  a solution to Eq.(\ref{CD_EVE4G_m}) if the following relations hold:
\be
c_{D-2} &=& -s~p(p + D - 3),
\eq{CD_EVG_c_D-2=}
\\
\l &=& -p(p + D - 2).
\eq{CD_EVG_l=}
\ee
Furthermore it is easy to show that there is another solution
to Eq.(\ref{CD_EVE4G_m}) with the same $\l$ and 
$c_{D-2}$, expressed as
\be
G'_{D-2} = S_1^p\int S_1^{-2p - D + 2} d\th_1,
\ee
where $p$ satisfies 
(\ref{CD_EVG_c_D-2=})
and
(\ref{CD_EVG_l=}).
Since 
$G_{D-2}$ and $G'_{D-2}$ are 
linearly independent, the general solutions to 
Eq.(\ref{CD_EVE4G_m}) are linear combinations of 
the above two special solutions.

From Eqs.
(\ref{CD_EVG_c_D-2=}),
(\ref{CD_EVG_l=}),
and
(\ref{CD_n_1=...=n_D-2})
we see
\be
&&
c_{D-2}=-s\ell(\ell + D - 3),
\sii
\ell \in \inte_{\ge 0},
\\
&&
\ell(\ell + D - 3)
=
p(p + D -3),
\eq{CD_ell-p}
\ee
{\it i.e.}
\be
(p - \ell)(p + \ell + D - 3) = 0.
\ee
Corresponding to the two solutions, $p=\ell$ and $p=-\ell - D + 3$, to the above 2nd order algebraic Eq. for $p$, 
the eigenvalues of ${\vDal}_{\rm LB}$ are
\be
\l = \kakkob -\ell(\ell + D -2), \sii\sii p = \ell \sii\sii \\
     -(\ell-1)(\ell + D - 3),\sii p = -\ell - D + 3
     \kakkoe, \sii \ell \in \inte_{\ge 0}.
\eq{CD_ell=}
\ee
There is no other eigenvalue. \vss

Let us show that $\l$ is further constrained by 
the requirement of the boundedness of $G_{D-2}$ and $G'_{D-2}$.
We divide the argument for the Euclid  and 
the Minkowski metric.\vss\vss

\noi (1) Euclid metric:

Since 
$S_1=\sin{\th_1}$
in this case,
the behavior of 
$G'_{D-2}$
around origin 
is 
$G'_{D-2} \sim \th_1^{-p - D + 3}$.
Corresponding to the solutions,
$p=\ell$ and $p=-\ell - D + 3$,
we have 
$G'_{D-2}\sim \th_1^{-\ell - D + 3}$
and 
$G'_{D-2} \sim \th_1^{\ell}$,
respectively.
From the boundedness at origin the former cases 
with $D\ge 4$ are excluded for all 
$\ell\in\inte_{\ge 0}$, 
and the latter cases are all permitted and have 
the eigenvalues 
$\l = -(\ell-1)(\ell + D - 3)$, 
as is seen by Eq.(\ref{CD_ell=}).

Since the behavior of 
the function 
$G_{D-2}=S_1^{p}$ 
at origin is
$G_{D-2}\sim \th_1^{p}$,
$p=\ell$ is permitted and have
the eigenvalue
$\l = -\ell(\ell + D -2)$.
The solutions with 
$p=-\ell - D + 3$ 
are permitted only in the case
$D\le 3 - \ell$, 
{\it i.e.},
the cases of $D=2,3$
\footnote{
Here we omit the trivial case , $D=1$.}.
After all the permitted eigenvalues are
\be
\l = -(\ell-1)(\ell + D - 3),
\sii \ell\in\inte_{\ge 0},
\sii ({\rm for}~{\rm Euclid}~{\rm metric}).
\eq{CD_eigenV_Eucli}
\ee
The eigenvalues 
$\l = -\ell(\ell + D - 2)$
are contained in (\ref{CD_eigenV_Eucli}).
In the case of $D=2$ the eigenvalues expressed in 
Eq.(\ref{CD_Euclid_l=}) are permitted, which are also 
contained in  (\ref{CD_eigenV_Eucli}).
Corresponding to these eigenvalues 
there are eigen-functions which are single-valued  and bounded at origin.

 \vss

\noi (2) Minkowski metric:

In this case there are three connected domains 
defined by 
(\ref{LB_defM+})
-
(\ref{LB_defM0}).
In $M_{\pm}$ we have $S_1 = \pm\sinh{\t}$, 
and
in $M_{0}$ we have $S_1 = -\cosh{\t}$.
Contrary to the case of Euclid metric 
we should require the boundedness condition 
at infinity ($\t\rarw\pm\infty$) not only at origin.

The eigen-functions should have the same eigenvalues 
in all domains, hence those functions 
blowing up in some domains are excluded.
The behavior of the functions under consideration 
is the same for the three domains, and the most singular domain is $M_{\pm}$, hence it is sufficient to consider there.

Firstly for the eigen-function 
$G'_{D-2}$
the behavior of it at origin and infinity 
is
$G'_{D-2}\sim e^{\t(-\ell - D + 2)}$
or
$G'_{D-2}\sim e^{\t(\ell - 1)}$
for the solutions
$p=\ell$ 
or
$p=-\ell - D + 3$, 
respectively.
Hence from the condition of 
the boudedness for the solution of $p=\ell$
we have 
$\ell = 2 - D$ and $\l = 0$.
For the solution of $p=-\ell - D + 3$
we have 
$\ell=1$
and also
$\l = 0$.
Concerning to 
$G_{D-2}=S_1^p$,  
from the boundedness 
for the solution of $p=\ell$,
we have $\ell=0$ and $\l=0$.
The solution of $p=-\ell - D + 3$ gives 
$\ell=3 - D$ and also $\l=0$.
In the case of $D=2$ 
the permitted eigenvalue is given 
by Eq.(\ref{CD_Minkow_l=}).
After all the eigenvalues in the Minkowski 
metric is expressed as
\be
\l = 
\kakkob 
~~0 , 
\sii\sii\si~ 
({\rm for}~D\ne 2,~{\rm Minkowski}~{\rm metric})
\\
L^2,\si L \in \real, 
\si~ ({\rm for}~D=2,~{\rm Minkowski}~{\rm metric})
\kakkoe.
\ee
Corresponding to these eigenvalues 
there are eigen-functions which are bounded
at origin and infinity.\vss

We have thus prepared to get the critical dimension of spacetime, through Eq.(\ref{Quanti_D=l}), {\it i.e.}, $D(D-4)/4=\l$.
In the case of Euclid metric  this gives
\be
\fr14 D(D-4) = -(\ell-1)(\ell + D - 3),
\sii \ell\in\inte_{\ge 0}
\ee
{\it i.e.},
\be
(D+2\ell - 6)(D + 2\ell - 2)=0.
\ee
Hence we get the solution
$
D=2,4,6,
$
for the Euclid metric.

In the case of Minkowski metric we have two cases, 
{\it i.e.}, $D=2$  with $\l\ge 0$, and the case of other $D$'s with $\l = 0$.
The former case is excluded since $D(D-4)/4= -1$ 
which does not coincide to $\l\ge 0$.
Thus we see the permitted case is only $D=4$ with 
$\l=0$, the main result of the present paper.
This result is a consequence of the 2nd-order 
secondary constraint, since in other dimension 
there exists no solution of the quantized equation of the constraint.
In that sense we can say that the critical dimension of the complex particle in the Minkowski spacetime is four.

 \vss


\section{Conclusion and speculations}

In the present paper we establish the fact that 
the first quantized theory of the complex particle 
in the Minkowski spacetime is consistent if and only if the dimension is four.
The result is different in the case of the Euclid spacetime, 
where $D=2,4$ and $6$ are permitted.
The classical theory of the model has some pathological features such as emergence of  the higher order secondary constraint, which is the key condition of deducing the critical dimension and is absent in the physically established models.
Such constraint may be related to the breakdown of Dirac's conjecture \cite{dirac_1},\cite{dirac_2}.
Concerning to the conjecture 
some authors have been studied in various contexts
 \cite{cawley_1}-\cite{kkk_2}, and there seems not firm agreement on it (see \cite{hori_1812},\cite{hori_1902} about the opinion of the author).

The next step is to build a field theory of the model, 
in order to promote the complex particle to a physically viable model.
A plausible candidate is the Chern-Simons type action  
proposed by Witten \cite{witten} in the string model.
It is expressed, for example, as
\be
{\cal I}
=
\int d\m~(	\ps\star Q\ps + g\ps\star\ps\star\ps), 
\ee
where $\star$ represents the so-called star-product,
satisfying some properties, $d\m$ is some measure and $Q$ is the BRST operator.
The nilpotency of $Q$ makes it possible to 
interpret $Q$ as something similar to 
the exterior derivative $d$ in the 
theory of the differential form.
If $\star$ and $d\m$ are properly defined 
the action may be invariant under
$
\d \ps = Q\L + g\ps\star\L,
$	
where $\L$ is infinitesimal parameter depending on 
the spacetime coordinates.
Required properties are the Leibniz rule, 
$
Q(\ps_1\star\ps_2) = (Q\ps_1)\star \ps_2 + \ps_1\star Q\ps_2
$, 
and vanishing of integral of {\it exact form},
$\int d\m~ Q\ps = 0$.
In the case of the complex particle 
we proposed a plausible candidate of $\star$ and 
$d\m$ in ref.\cite{hori_05}.
The Leibniz rule is easily satisfied but the 
vanishing of $\int d\m~ Q\ps$ restricts the spacetime dimension.
Surprisingly, the above requirement leads to 
$D=4$. 
However, the field theory of the complex particle 
is under construction at present because of some problems to be overcome.

The string theories 
 in 10 or 26 dimensions inevitably have extra dimensions which brings on the model with Calabi-Yau  manifold \cite{hitchin} or the theory of landscape \cite{susskind}.
They may incorporate a quantum theory of gravity, but there exists absolutely no experimental evidence for the phenomenology of them. 
Although the present stage of the complex particle is a toy model, there may be possibility that it provides an alternative starting line to a realistic model without extra dimensions.

\renewcommand{\thesection}{\Alph{section}}
\setcounter{section}{0}
\setcounter{equation}{0}
\makeatletter
  \renewcommand{\theequation}{%
     \thesection.\arabic{equation}}
  \@addtoreset{equation}{section}
\makeatother
\setcounter{section}{0}
\setcounter{equation}{0}	
\section{\!\!\!\!\!\!ppendix}

\begin{xx}\eq{App_dL_xx}

{\rm 

 \si Under the transformation of $\bbz$ and $g$, 
\be
\d \bbz = \ve \bbw + \fr{\ve_0}{\bar{g}}\bar{\bbw},
\sii
\d g = \x,
\eq{App_d_sg}
\ee
the lagrangian of the complex particle varies as
\be
\d L
&=&
\fr{\ell_0}{2g\bar{g}}
[\dot{\ve}_0 - 2i\k g\bar{g}(\ve - \bar{\ve})]
+
A\ell + \fr{dE}{d\t}
+
{\rm c.c.},
\eq{App_dL_cplx}
\ee
where
$A$ and $E$ are defined by 
Eqs.(\ref{CPLX_def_A_E}).\\
\noi\sen

\noi pr.)

Starting with the expression
\be
	\d L
&=&
\fr{1}{g}\bbw\cdot
\left(
\ve \bbw + \fr{\ve_0}{\bar{g}}\bar{\bbw}
\right)^\bullet
-
\fr{\bbw^2}{2g^2}\x
+
i\k
\left[
\left(
\ve \bbw + \fr{\ve_0}{\bar{g}}\bar{\bbw}
\right)^\bullet\cdot\bar{\bbz}
+
\bbw\cdot
\left(
\bar{\ve}\bar{\bbw}
+
\fr{\ve_0}{g}\bbw
\right)
\right]
+
{\rm c.c.}
\nn
&=&
\ve
\left\{
\und{\fr{(\bbw^2)^\bullet}{2g}}
+
i\k
(
\dot{\bbw}\cdot\bar{\bbz}
-
\bbw\cdot\bar{\bbw}
)
\right\}
+
\dot{\ve}
\left(
\und{\fr{\bbw^2}{g}}
+
i\k
\bbw\cdot\bar{\bbz}
\right)
-
\fr{\bbw^2}{2g^2}\x
\nn
&& \sii
+
\ve_0
\left\{
\left(
\uwave{\fr{\bbw}{g}}
+
i\k\bar{\bbz}
\right)
\cdot\left(\fr{\bar{\bbw}}{\bar{g}}\right)^\bullet
+
i\k
\fr{1}{g}\bbw^2
\right\}
+
\dot{\ve}_0
\left\{
\fr{\bbw\cdot\bar{\bbw}}{g\bar{g}}
+
i\k\fr{\bar{\bbw}\cdot\bar{\bbz}}{\bar{g}}
\right\}
+
{\rm c.c.},
\eq{CPLX_1xx}
\ee
the terms of total derivatives are picked up 
by using the following relations:
\be
\ve\fr{(\bbw^2)^\bullet}{2g}
+
\dot{\ve}\fr{\bbw^2}{g}
&=&
\fr{\bbw^2}{2g^2}(\ve g)^\bullet
+
\fr{d}{d\t}
\left(
\ve\fr{\bbw^2}{2g}
\right),
\\
\ve(\dot{\bbw}\cdot\bar{\bbz} - \bbw\cdot\bar{\bbw})
+
\dot{\ve}\bbw\cdot{\bar{\bbz}}
&=&
-2\ve\bbw\cdot\bar{\bbw} 
+ 
\fr{d}{d\t}
(\ve\bbw\cdot\bar{\bbz}),
\\
\ve_0\fr{\bbw}{g}\cdot
\left(\fr{\bar{\bbw}}{\bar{g}}\right)^\bullet
+
{\rm c.c.}
&=&
-
\fr{\dot{\ve}_0}{2}
\fr{\bbw\cdot\bar{\bbw}}{g\bar{g}}
+
\fr{d}{d\t}
\left(
\ve_0\fr{\bbw\cdot\bar{\bbw}}{2g\bar{g}}
\right)
+
{\rm c.c.},
\\
\ve_0\bar{\bbz}\cdot
\left(
\fr{\bar{\bbw}}{\bar{g}}
\right)^\bullet
+
\dot{\ve}_0\fr{\bar{\bbz}\cdot\bar{\bbw}}{\bar{g}}
&=&
-
\ve_0\fr{\bar{\bbw}^2}{\bar{g}}
+
\fr{d}{d\t}
\left(
\ve_0
\fr{\bar{\bbz}\cdot\bar{\bbw}}{\bar{g}}
\right).
\ee
Then
\be
\d L
&=&
\fr{\bbw\cdot\bar{\bbw}}{2g\bar{g}}
(\dot{\ve}_0 - 4i\ve\k g\bar{g})
+
((\ve g)^\bullet + 4i\k g\ve_0 - \x)
\fr{\bbw^2}{2g^2}
\nn
&&\sii
+
\fr{d}{d\t}
\left\{
\ve
\left(
\fr{\bbw^2}{2g} + i\k\bbw\cdot\bar{\bbz}
\right)
+
\ve_0
\left(
i\k\fr{\bar{\bbw}\cdot\bar{\bbz}}{\bar{g}}
+
\fr{\bbw\cdot\bar{\bbw}}{2g\bar{g}}
\right)
\right\}
+
{\rm c.c.},
\ee
where we used 
$
i(\bar{\bbw})^2/\bar{g} + {\rm c.c.}
=
-i\bbw^2/g + {\rm c.c.}
$.
Writing the term of c.c. explicitly, we obtain 
Eq.(\ref{App_dL_cplx}).
  \qed\\
}
\sen

\end{xx}


 \vss
 
\begin{xx}\eq{App_xx_dal-vDal}
\si 
{\rm 
The d'Aembert and the Laplace-Beltrami operator 
are related by}
\be
s~\dal_u 
=
\fr{1}{r^2}{\vDal}_{\rm LB} 
+ \fr{1}{r^{D-1}}\pdel{}{r} r^{D-1}\pdel{}{r} , 
\sii
s \deff {\rm sgn}(u_{\m}u^\m).
\eq{App_dal-vDal}
\ee
\noi\sen

\noi pr.)

From the definition of ${\vDal}_{\rm LB}$ we have
\be
{\vDal}_{\rm LB}
&=&
\fr12
\left(
u^\m\pdel{}{u_\n} - u^\n\pdel{}{u_\m}
\right)
\left(
u_\m\pdel{}{u^\n} - u_\n\pdel{}{u^\m}
\right)
\nn
&=&
u_\m u^\m\dal_u
-
(D-1)\tilde{r}D_r
-
u^\m u^\n\fr{\del^2}{\del u^\m \del u^\n} ,
\eq{App_1}
\ee
where
\be
D_r \deff \fr{u^\m}{\tilde{r}}\pdel{}{u^\m} ,
\sii
\tilde{r} \deff \sqrt{s~u_\m u^\m}.
\ee
Writting $D_r^2$ as
\be
D_r^2
&=&
\fr{u^{\m}}{\tilde{r}}
\left(	
\pdel{}{u^{\m}} \fr{u^\n}{\tilde{r}}
\right)\pdel{}{u^\n}
+
\fr{u^{\m}u^\n}{\tilde{r}^2}\ppdel{}{u^\n}{u^{\m}} ,
\ee
we see the 1st term in r.h.s. of the above eqation 
is
\be
({\rm 1st}~{\rm term}~{\rm in}~{\rm r.h.s.})
=
\fr{u^\m}{\tilde{r}}
\left(
\fr{\d^{\n}_{\m}}{\tilde{r}}
-
\fr{s~u^\n u_\m}{\tilde{r}^3}
\right)\pdel{}{u^\n}
=
0.
\ee
Hence
\be
D_r^2
=
\fr{u^{\m}u^{\n}}{\tilde{r}^2}\ppdel{}{u^\n}{u^{\m}} ,
\eq{app_f16}
\ee
and from 
Eq.(\ref{App_1})
we get
\be
{\vDal}_{\rm LB}
=
s\tilde{r}^2\dal_u
-
(D-1)\tilde{r}D_r
-
\tilde{r}^2D_r^2.
\ee
Using 
Eq.(\ref{LB_f6}) 
we get
Eq.(\ref{App_dal-vDal}).  \qed

\noi\sen

\end{xx}


\end{document}